\documentstyle[aps,prb,psfig,floats]{revtex}
\begin{document}
\draft
\twocolumn[\hsize\textwidth\columnwidth\hsize
\csname @twocolumnfalse\endcsname
\title{
Effect of a Normal-State Pseudogap on Optical Conductivity
in Underdoped Cuprate Superconductors
}
\author{T. Dahm}
\address{Max-Planck-Institut f\"ur Physik komplexer Systeme,
N\"othnitzer Str. 38, D-01187 Dresden, Germany}
\author{D. Manske}
\address{Institut f\"ur Theoretische Physik, Freie Universit\"at Berlin,
Arnimallee 14, D-14195 Berlin, Germany}
\author{L. Tewordt}
\address{I. Institut f\"ur Theoretische Physik,
Universit\"at Hamburg, Jungiusstr. 9, D-20355 Hamburg,
Germany}
\date{\today}
\maketitle
\begin{abstract}
We calculate the c-axis infrared conductivity $\sigma_c(\omega)$
in underdoped
cuprate superconductors for spinfluctuation exchange scattering
within the
CuO$_2$-planes including a phenomenological $d$-wave pseudogap
of amplitude
$E_g$. For temperatures decreasing below a temperature
$T^* \sim E_g/2$, a gap
for $\omega < 2E_g$ develops in $\sigma_c(\omega)$ in the incoherent
(diffuse) transmission limit. The resistivity shows 'semiconducting'
behavior, i.e. it increases for low temperatures above the
constant behavior
for $E_g=0$. We find that the pseudogap structure in the
in-plane optical
conductivity is about twice as big as in the interplane conductivity 
$\sigma_c(\omega)$, in qualitative agreement with experiment.
This is a
consequence of the fact that the spinfluctuation exchange
interaction is
suppressed at low frequencies as a result of the opening of
the pseudogap.
While the c-axis conductivity in the underdoped regime is
described best
by incoherent transmission, in the overdoped regime coherent
conductance
gives a better description.
\end{abstract}
\pacs{74.25.Gz, 74.20.Mn, 74.72.-h, 72.15.-v}
]
\narrowtext

%
%
Numerous experiments have established the fact that the underdoped
cuprate superconductors exhibit a 'pseudogap' behavior in both
spin and charge degrees of freedom below a characteristic
temperature $T^*$
which can be well above the superconducting transition
temperature $T_c$.
Many interpretations of the pseudogap have been advanced
(see, e.g. the
discussion in Ref. \onlinecite{Williams}), however, no
consensus has been 
reached so far, which of the various microscopic theories
is the correct one.
It has been shown by Williams et al. \cite{Williams} that
specific heat,
susceptibility and NMR data of many underdoped cuprates
can successfully
be modeled using a phenomenological normal-state pseudogap having
$d$-wave symmetry and an amplitude which is temperature independent
but increases upon lowering the doping level into the
underdoped regime.
This strong anisotropy of the pseudogap is also in accordance with
angle-resolved photoemission spectroscopy (ARPES) experiments on
underdoped Bi$_2$Sr$_2$CaCu$_2$O$_{8-\delta}$ (Bi2212) 
\cite{Harris,Norman}. This model yields a smooth evolution of the
normal state pseudogap into the superconducting gap as has been
found in scanning-tunneling (STM) experiments \cite{Renner}. Also,
measurements of resistivity, Hall coefficient and thermoelectric
power can be reconciled with this model \cite{Cooper,Batlogg}.
However, measurements of the dynamical conductivity within or
perpendicular to the CuO$_2$-planes consistently show pseudogap
structures having a size which only weakly depends on doping in the
underdoped regime, in marked contrast to the model above. In addition,
while the c-axis conductivity in YBa$_2$Cu$_3$O$_{7-\delta}$ (YBCO)
shows a pseudogap having a size of approximately 300-400 cm$^{-1}$, 
\cite{Homes,Uchida} the size of the pseudogap extracted from 
ab-plane conductivity is of the order of 600-700 cm$^{-1}$.
\cite{Puchkov,Basov} This difference cannot be attributed to the
charge reservoir layers between the CuO$_2$-planes, as it has been
convincingly shown recently that the pseudogap seen in c-axis
conductivity has its origin in the CuO$_2$-planes \cite{Bernhard}.

Here, we try to address these apparent inconsistencies in an effort
to come to a consistent phenomenological description of the pseudogap.
We calculate the c-axis and ab-plane conductivity in the presence of
a temperature independent, but doping dependent pseudogap
as is suggested
by the work of Williams et al \cite{Williams}. To do so it is necessary
to take into account some scattering mechanism within the CuO$_2$-plane.
We choose to study spinfluctuation exchange scattering within the
so-called selfconsistent fluctuation-exchange (FLEX) approximation for
the two-dimensional Hubbard model \cite{Bickers}, which has proven to
give a good qualitative description of the high-$T_c$ cuprates in the
optimally doped regime \cite{Bickers,Pao,Monthoux,DahmTewordt}.
Especially, the selfconsistently calculated interaction due to
exchange of spin and charge fluctuations yields a quasiparticle
scattering rate which varies linearly with frequency in the
normal state,
and exhibits a gap-like suppression at lower frequencies in
the superconducting state. At the same time the effective mass ratio
is increased at lower frequencies in the superconducting state. Thus, 
FLEX approximation
accounts for the damping and mass enhancement needed to extend
the theory
for thermodynamical quantities by Williams et al. \cite{Williams} to
dynamical quantities.

We have shown previously that such a theory is capable of describing
qualitatively a number of different experiments in underdoped cuprates
like Knight-shift, nuclear-spin relaxation rate, neutron scattering,
ARPES, STM,
and ab-plane conductivity experiments \cite{DMT2,DMTRaman}.
In particular, the pseudogap leads to a gap-like suppression of the
scattering rate at lower frequencies below the linear frequency
extrapolation in accordance with the observed in-plane conductivity. 
The $T_c$ which is calculated selfconsistently
from the strong-coupling gap equation is suppressed in proportion
to the magnitude of the pseudogap. The behavior of the Knight-shift, 
nuclear-spin relaxation rate, and density of states is described
correctly as the temperature is decreased through $T_c$, showing
a smooth evolution of the pseudogap into the superconducting gap.

Here, we will show that the selfconsistency of the spinfluctuation
interaction with the single particle properties, especially with
the pseudogap itself, provided by the FLEX approximation will lead
to a natural understanding of the differences in size of
the pseudogap
seen in ab-plane and c-axis conductivity. Also, the
semiconducting behavior
of the c-axis resistivity can be understood qualitatively 
\cite{Homes,Uchida,Prelovsek}. However, we will also see that such a 
simple ansatz for the pseudogap is not sufficient to understand the
doping independence of the size of the gap observed in ab-plane
and c-axis conductivity.

Measurements of the c-axis conductivity suggest that conductance in
c-direction is coherent in the overdoped regime \cite{Bernhard}
successively becoming incoherent in the underdoped regime 
\cite{Homes,Uchida}. We will therefore study the two limits of
coherent and incoherent c-axis conductivity. We will see that indeed
within our model incoherent conductance gives a good description of
the underdoped regime, while coherent conductance is more appropriate
for the overdoped regime, confirming previous interpretations of
the c-axis conductivity.

The coherent conductivity along the interplane c-direction is
given to lowest order in the inter-layer hopping $t_\perp$
\cite{Prelovsek} by
\begin{eqnarray}
\lefteqn{
\sigma_c \left( \omega \right) = \frac{e^2 t_\perp^2 c_0}{\hbar a_0^2} 
\frac{\pi}{\omega} \int_{-\infty}^{\infty} d\omega'
\left[f(\omega') - f(\omega'+\omega)\right]
}\nonumber\\
& \times &
\frac{1}{N} \sum_{\bf k}
\left[N({\bf k},\omega'+\omega)N({\bf k},\omega')
\right.
\nonumber\\
& + &
\left.
A_1({\bf k},\omega'+\omega)A_1({\bf k},\omega') +
A_g({\bf k},\omega'+\omega)A_g({\bf k},\omega')
\right],
\label{eq1}
\end{eqnarray}
where $c_0$ and $a_0$ are the c-axis and ab-plane lattice constants.
Here, $N$ is the normal spectral function and $A_1$ and $A_g$ are the
anomalous spectral functions with respect to the superconducting gap
and the pseudogap, respectively. These spectral functions are taken
from a selfconsistent solution of the FLEX equations in the presence
of the pseudogap and are given by:
\begin{eqnarray}
N({\bf k},\omega) & = & - \frac{1}{\pi} {\rm Im} 
\frac{\omega Z + \epsilon_k
+ \xi}{\left(\omega Z \right)^2 - \left(
\epsilon_k+ \xi \right)^2 -
E_g^2 - \phi^2}
\label{eq2} \\
A_1({\bf k},\omega) & = & - \frac{1}{\pi} {\rm Im} \frac{\phi}
{\left(\omega Z \right)^2 - \left( \epsilon_k+ \xi \right)^2 -
E_g^2 - \phi^2}
\label{eq3} \\
A_g({\bf k},\omega) & = & - \frac{1}{\pi} {\rm Im} \frac{E_g}
{\left(\omega Z \right)^2 - \left( \epsilon_k+ \xi \right)^2 -
E_g^2 - \phi^2}
\label{eq4} 
\end{eqnarray}
\begin{figure}[t]
\vspace{-1.0cm}
\centerline{\psfig{file=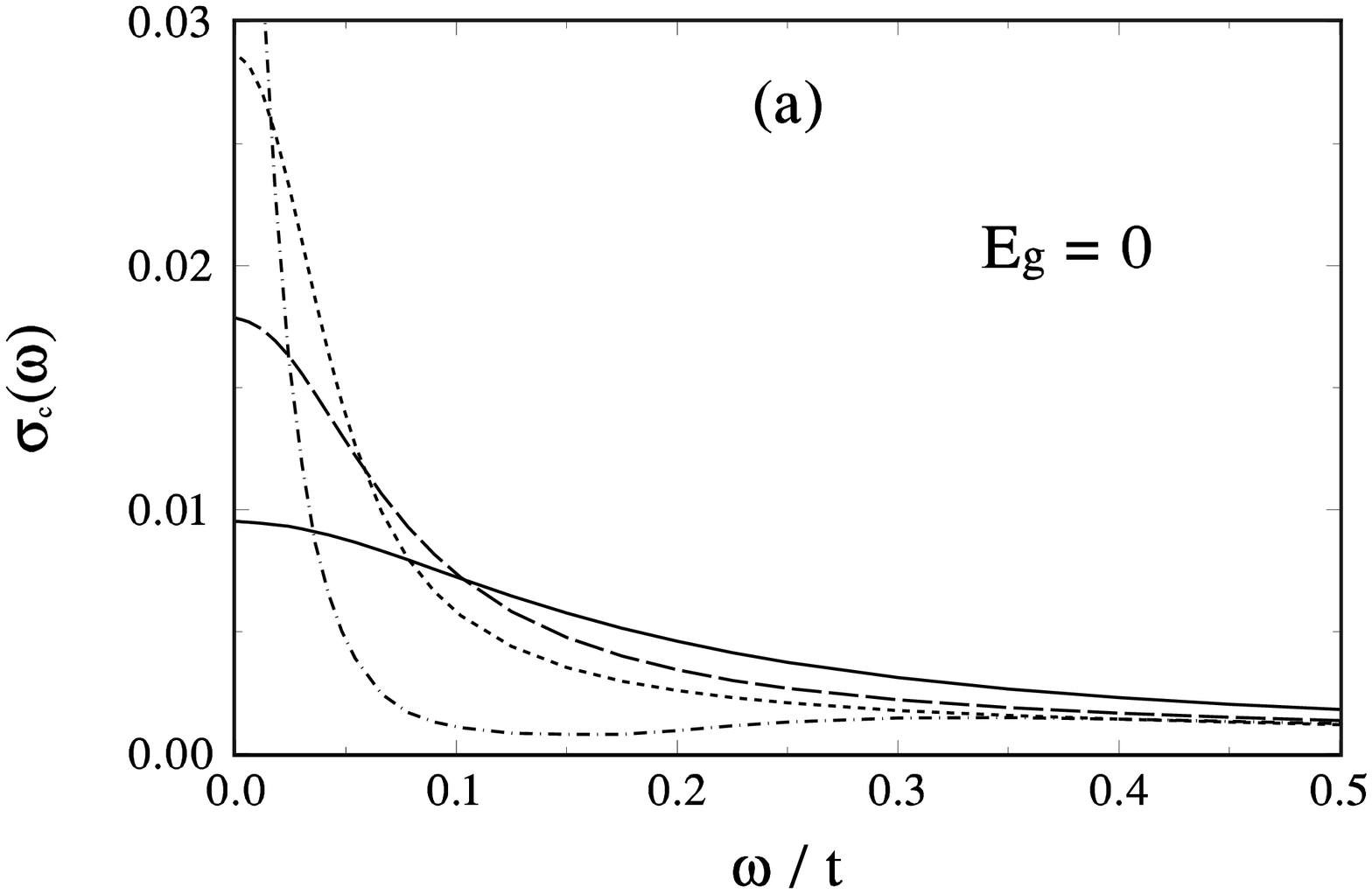,height=2.8in,angle=-90}}
\vspace{-1.0cm}
\centerline{\psfig{file=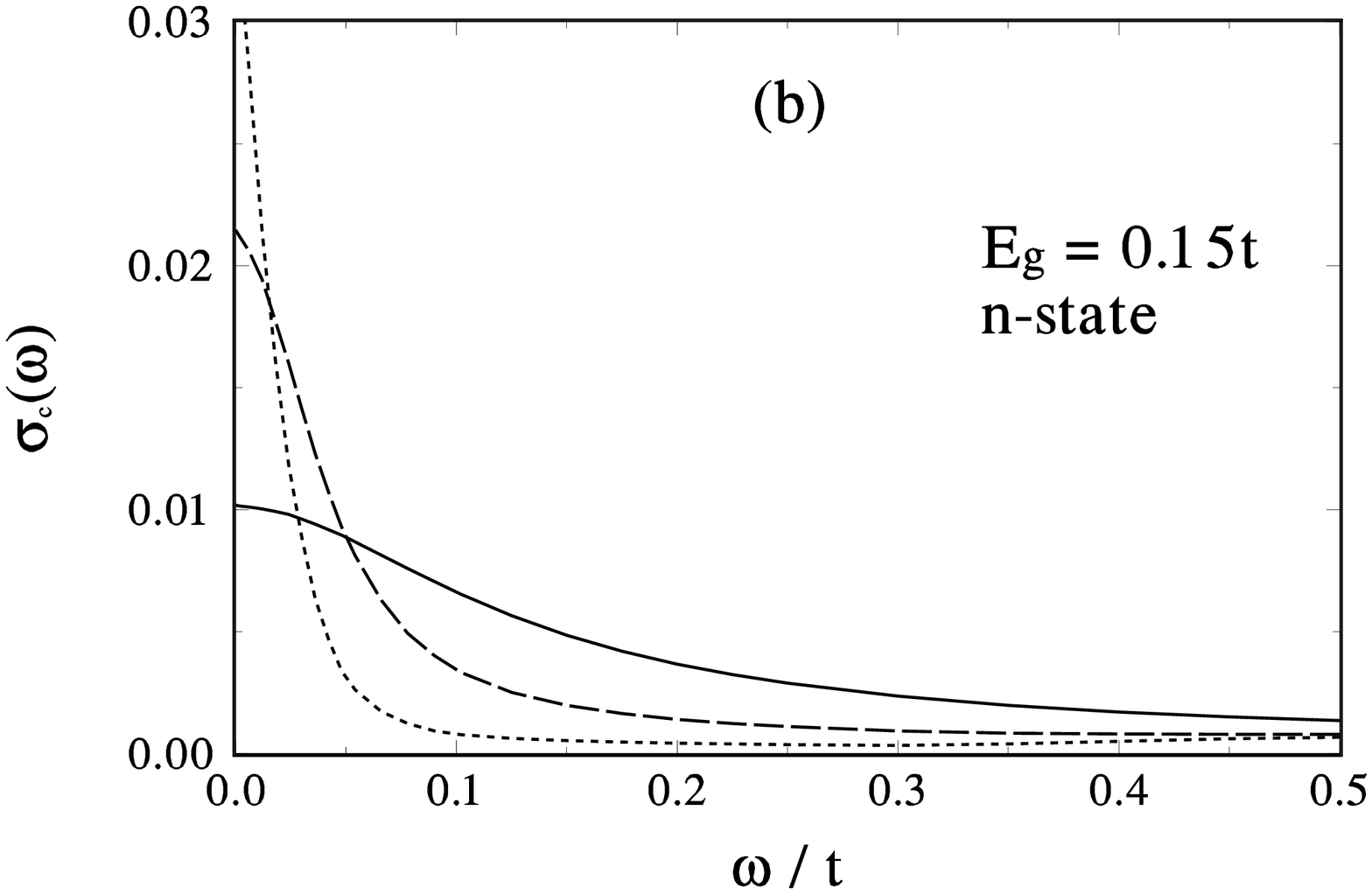,height=2.8in,angle=-90}}
%
\caption{The coherent dynamical c-axis conductivity (Eq. (\ref{eq1}))
for three temperatures $T$=0.1$t$ (solid line), 0.05$t$ (dashed), 
and 0.03$t$ (dotted) (a) for a pseudogap amplitude $E_g=0$ 
(b) for $E_g=0.15t$.
The dashed-dotted line in (a) holds for the superconducting
state for $E_g=0$ ($T_c=0.023t$) at $T=0.017t$.}
\label{fig1}
\end{figure}
We want to emphasize that it is necessary to include the bubble
contribution due to $A_g$ into the conductivities and 
susceptibilities. Neglection of this term leads to severe 
disagreement with the data.
Following Ref. \onlinecite{Williams}, for the pseudogap we assume
the form
\begin{equation}
E_g\left( {\bf k} \right) = E_g \left[ \cos k_x - \cos k_y \right]
\label{eq5} 
\end{equation}
where $E_g$ is temperature independent and increases with decreasing
doping level below the optimal doping level. The FLEX equations are
solved selfconsistently in the presence of this pseudogap and yield
the quasiparticle selfenergy components $Z({\bf k},\omega)$ and
$\xi({\bf k},\omega)$ as well as the superconducting gap
$\phi({\bf k},\omega)$. Within FLEX the effective spin and
charge fluctuation interactions are given by the RPA expressions
\begin{equation}
\frac{3}{2} U^2 \frac{\chi_{s0}}{1-U \chi_{s0}} \qquad {\rm and}
\qquad \frac{1}{2} U^2 \frac{\chi_{c0}}{1+U \chi_{c0}} ,
\label{eq6} 
\end{equation}
where the bubble spin susceptibility $\chi_{s0}$ is calculated
selfconsistently from the expression
\begin{eqnarray}
\lefteqn{
{\rm Im} \chi_{s0}({\bf q},\omega)
=
\pi\int_{-\infty}^{\infty}d\omega'
\left[f(\omega') - f(\omega'+\omega)\right]
}\nonumber\\
& \times &
\frac{1}{N} \sum_{\bf k}
\left[N({\bf k}+{\bf q},\omega'+\omega)N({\bf k},\omega') +
A_1({\bf k}+{\bf q},\omega'+\omega)
\right.
\nonumber\\
& \times &
\left.
A_1({\bf k},\omega') +
A_g({\bf k}+{\bf q},\omega'+\omega)A_g({\bf k},\omega')
\right]
\label{eq7}
\end{eqnarray}
and thus depends on the pseudogap via the spectral functions.
The charge susceptibility $\chi_{c0}$ has the opposite sign
in front of the anomalous terms in Eq. (\ref{eq7}). For the
on-site Coulomb repulsion we take an effective $U({\bf q})$
with maximum value $U=3.6$ at ${\bf q} = {\bf Q} = (\pi,\pi)$,
as has been discussed in Ref. \onlinecite{DMTRaman}.
For the bandstructure $\epsilon_k$ we take the tight-binding
band
\begin{equation}
\epsilon_k = t \left[ -2 \cos k_x - 2 \cos k_y + 4 B \cos k_x
   \cos k_y - \mu \right]
\label{eq8} 
\end{equation}
where $t$ is the in-plane hopping matrix element and we take
$B=0.45$ and $\mu=-1.1$ which describes approximately the
Fermi surfaces of the YBCO and Bi2212 compounds.

In Fig. \ref{fig1}(a) we show the coherent conductivity
$\sigma_c ( \omega )$ calculated in this way from Eq. (\ref{eq1})
for different temperatures. Here we took $E_g = 0$. For
decreasing temperatures $T$ a coherent Drude peak develops at low
frequencies. Such a development of a coherent Drude peak has been
observed in overdoped cuprates \cite{Bernhard,Homes}, where the
pseudogap is absent or small. Thus, our coherent conductance results
account well for this observation in the overdoped compounds.
In the superconducting state a suppression of $\sigma_c ( \omega )$
at intermediate frequencies sets in as shown by the
dashed-dotted line in Fig. \ref{fig1}(a) for $T=0.017t$. Here,
$T_c=0.023t$. At the same time the Drude peak continues to
sharpen.
Fig. \ref{fig1}(b) shows the normal-state coherent conductivity
$\sigma_c ( \omega )$ 
in the presence of the pseudogap with amplitude
$E_g =0.15 t$. While
the pseudogap leads to a suppression of $\sigma_c ( \omega )$ at
intermediate frequencies, the coherent Drude peak at low frequencies
still remains and even sharpens, similar to the superconducting
state in Fig. \ref{fig1}(a). These results are
completely different from the experimental results in the normal 
state of underdoped cuprates, showing instead of a coherent Drude
peak a gap-like suppression at low frequencies. Thus, it is not
sufficient to simply turn on a pseudogap in order to account for the
c-axis conductivity in underdoped compounds. As has been noted earlier
\cite{Uchida,Hirschfeld} the c-axis conductance at the
same time becomes
incoherent in the underdoped regime and therefore it is necessary to
calculate the incoherent conductivity in the presence of a pseudogap.
Incoherent conductivity corresponds to diffuse c-axis transmission
and amounts in taking the averages of the spectral functions
$N({\bf k},\omega)$, $A_1({\bf k},\omega)$, and
$A_g({\bf k},\omega)$
over all momenta (see the discussion in Ref.
\onlinecite{Hirschfeld}). 
This means that $N({\bf k},\omega)$ is replaced by
the density of states 
\begin{equation}
 N \left( \omega \right) = \frac{1}{N} \sum_k N({\bf k},\omega)
\label{eq9} 
\end{equation}
while the averages of $A_1$ and $A_g$ vanish due to the $d$-wave
symmetry of the superconducting and the pseudogap \cite{Hirschfeld}. 
Then we find:
\begin{eqnarray}
\sigma_c^{\rm incoh} \left( \omega \right) & = & 
\frac{e^2 t_\perp^2 c_0}{\hbar a_0^2} 
\frac{\pi}{\omega} \int_{-\infty}^{\infty} d\omega'
\left[f(\omega') - f(\omega'+\omega)\right]
\nonumber\\
& & \times 
N(\omega'+\omega)N(\omega') .
\label{eq10}
\end{eqnarray}

\begin{figure}[ht]
\vspace{-1.0cm}
\centerline{\psfig{file=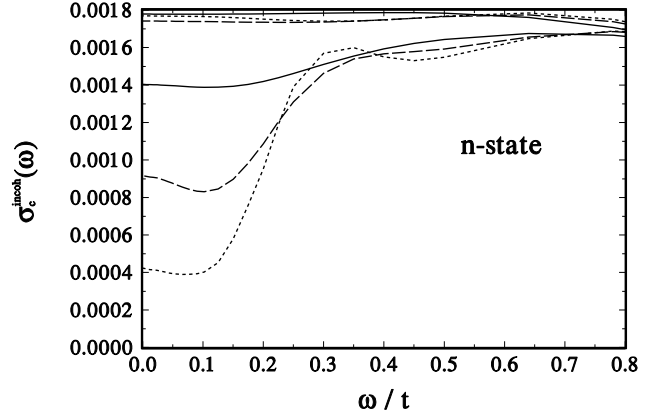,height=2.8in,angle=-90}}
%
\caption{The incoherent dynamical c-axis conductivity
(Eq. (\ref{eq10}))
for three temperatures $T$=0.1$t$ (solid line), 0.05$t$ (dashed), 
and 0.03$t$ (dotted) and two values of
$E_g$=0 (upper three curves) and 0.15$t$ (lower three curves).
}
\label{fig2}
\end{figure}

Fig. \ref{fig2} shows the incoherent conductivity 
$\sigma_c^{\rm incoh} ( \omega )$ for $E_g=0$ and $E_g =0.15 t$
for three different temperatures $T$ = 0.1, 0.05, and 0.03$t$.
For $E_g =0.15 t$ a gap develops below a threshold frequency
of about $\omega \sim 2 \tilde{E}_g$ upon lowering the temperature 
while the conductivity stays almost
constant for frequencies above this thres\-hold energy.
Here, $\tilde{E}_g =
2 E_g / {\rm Re} Z(E_g)$ is the renormalized amplitude of
the $d$-wave
pseudogap, and Re$Z(\omega)$ is the average mass
renormalization at
the Fermi surface which is of the order of 2 for the 
parameters considered here. For $E_g=0$, $\sigma_c^{\rm incoh}$
is very 
much frequency and temperature independent.
These results are in qualitative agreement with the
measured interplane
conductivity in underdoped YBCO compounds \cite{Homes}.
The gap in the c-axis conductivity
$\sigma_c^{\rm incoh} ( \omega )$
for frequencies $\omega$ below $2 \tilde{E}_g$ develops below a 
characteristic temperature $T^* \sim \tilde{E}_g/2$.
We want to stress that the temperature evolution of all physical
quantities arises exclusively from the Fermi and Bose functions
occuring in the FLEX equations in our real-frequency formulation
\cite{DahmTewordt} and in the expressions for the susceptibilities
(see Eq. (\ref{eq7})), since we assumed that
the pseudogap $E_g({\bf k})$
defined in Eq. (\ref{eq5}) is temperature independent, following Ref.
\onlinecite{Williams}. Physically, this means that
above $T^*$ the effect of
the pseudogap is smeared out such that the normal state behavior
(corresponding to the FLEX equations
for $E_g=0$) is recovered while the effect of
the pseudogap on the quasiparticle and spin excitation spectra
increases as $T$ decreases below $T^*$ towards $T_c$.
\begin{figure}[ht]
\vspace{-1.0cm}
\centerline{\psfig{file=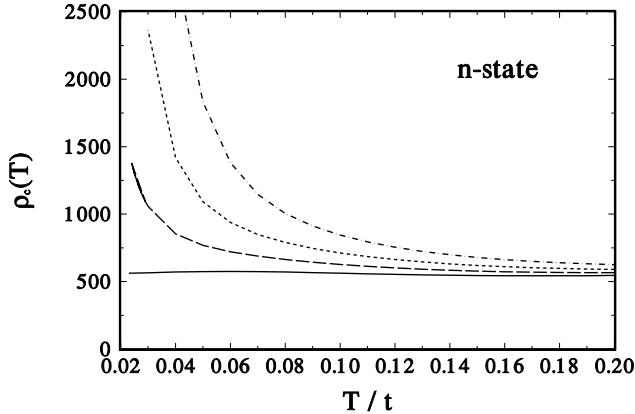,height=2.8in,angle=-90}}
%
\caption{The incoherent c-axis resistivity $\rho_c$ as a function
of temperature for $E_g$=0 (solid line), 0.1$t$ (dashed), 0.15$t$
(dotted), and 0.2$t$ (dashed-dotted).
}
\label{fig3}
\end{figure}

From our incoherent conductivity we can extract the
c-axis resistivity
$\rho_c = [ \sigma_c^{\rm incoh} ( \omega=0 ) ]^{-1}$
in the presence of
the pseudogap. The temperature dependence of $\rho_c$ is shown in
Fig. \ref{fig3} for different values of the pseudogap
amplitude $E_g$=0, 0.1, 0.15, and 0.2$t$, corresponding to
different doping levels. For $E_g=0$, $\rho_c$ is almost
constant. For
finite $E_g$ it starts to increase above the curve for
$E_g=0$ at lower 
temperatures. This 'semiconducting' behavior of $\rho_c$
is directly
related to the depth of the pseudogap at zero frequency in
$\sigma_c^{\rm incoh} ( \omega=0 )$ which increases for
increasing $E_g$
and decreasing temperature (see Fig. \ref{fig2}).
The results in Fig.
\ref{fig3} are consistent with the estimate of the
characteristic temperature
given above, i.e.
$T^* \sim \tilde{E}_g/2 = E_g / {\rm Re} Z$, because
the steep rise of $\rho_c$ for a given $E_g$
appears approximately
below $T^*$. Our results are in qualitative agreement with c-axis 
resistivity data on underdoped YBCO \cite{Homes,Uchida}. We remark
that the definition of the characteristic temperature
$T^* \sim \tilde{E}_g/2$
corresponds to the scaling procedure in Ref.
\onlinecite{Williams} where
it has been shown that the NMR Knight shift
$ ^{89}K_n(T)$ for a wide
range of doping values follows closely a universal
scaling curve if the
data are plotted against a scaling parameter $z=2T/\tilde{E}_g$.
The downturn of $ ^{89}K_n(T)$ for decreasing $z$
occurs at about $z=1$
which corresponds to a $T^* \sim \tilde{E}_g/2$.

\begin{figure}[t]
\vspace{-1.0cm}
\centerline{\psfig{file=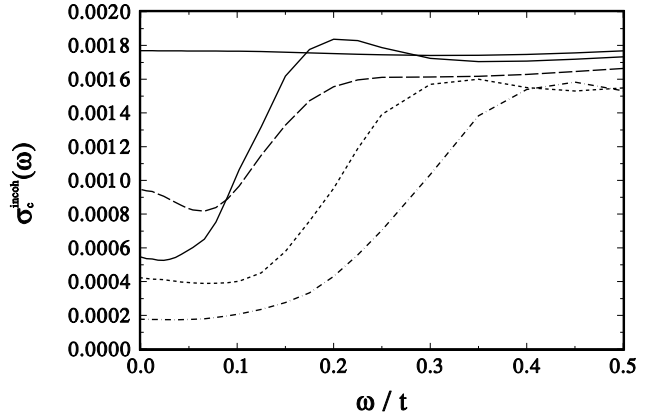,height=2.8in,angle=-90}}
%
\caption{The incoherent dynamical c-axis conductivity (Eq. (\ref{eq10}))
for $T=0.03t$ and $E_g$=0 (solid line), 0.1$t$ (dashed),
0.15$t$ (dotted), and 0.2$t$ (dashed-dotted).
For comparison we also show the result in the superconducting
state for $E_g=0$ ($T_c=0.023t$) at $T=0.017t$ (lower solid line).
}
\label{fig4}
\end{figure}

In Fig. \ref{fig4} we show $\sigma_c^{\rm incoh} ( \omega )$ at a 
fixed temperature $T=0.03t$ and different values of $E_g$. 
From Fig. \ref{fig4} we see that the
renormalized size $2 \tilde{E}_g$ of the gap in $\sigma_c^{\rm incoh}$
follows $E_g$. If one assumes that $E_g$ strongly changes with
doping level, as has been proposed in Ref. \onlinecite{Williams},
the results shown in Fig. \ref{fig4} cannot provide an explanation
for the doping independence of the pseudogap seen in c-axis
conductivity on underdoped cuprates. This is an apparent inconsistency
of the model by Williams et al, which can describe well thermodynamic
quantities, but does not give a satisfactory account of the doping 
dependence of ab-plane and c-axis conductivity, which are dynamical 
quantities. Our results rather indicate
that there are two independent energy scales involved in the pseudogap 
problem:
first, the width of the pseudogap (here $2\tilde{E}_g$)
as it is seen in the $\omega$-dependence of the c-axis
and ab-plane conductivity, being largely doping independent, and
second, the depth of the pseudogap as it is seen in the $\omega=0$
value of the incoherent c-axis conductivity corresponding to the
characteristic temperature $T^*$ 
for the c-axis resistivity and the thermodynamic quantities,
which increases upon lowering the doping level.
Such two energy
scales could be introduced into the problem by considering more
complicated forms for the pseudogap than Eq. (\ref{eq5}). For example,
the pseudogap could have a frequency dependence or a momentum
dependence which changes with temperature as suggested by a recent
analysis of ARPES data \cite{Norman,Williams}.

The lower solid line in Fig. \ref{fig4} shows the result for
the incoherent c-axis conductivity in the superconducting state
for $T=0.017t$ and $E_g=0$ ($T_c=0.023t$). This compares well
with the experimental results on optimally doped YBCO, \cite{Homes}
showing a suppression at low frequencies, similar to the
suppression due to the normal-state pseudogap. In addition,
a weak enhancement at the gap edge develops. Again, this behavior
is completely different from the corresponding behavior of the
coherent c-axis conductivity (see the dashed-dotted line in 
Fig. \ref{fig1}(a) ).

\begin{figure}[t]
\vspace{-1.0cm}
\centerline{\psfig{file=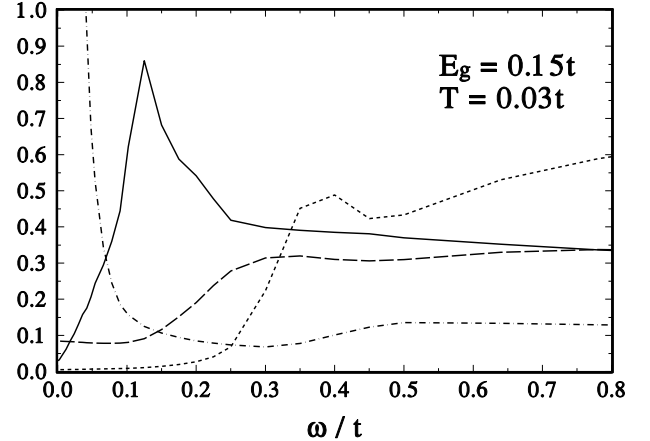,height=3.0in,angle=-90}}
%
\caption{Density of states $N(\omega)$ (solid line), incoherent
c-axis conductivity $\sigma_c^{\rm incoh} (\omega)$ (dashed), 
quasiparticle damping rate $\omega {\rm Im} Z({\bf k_a},\omega)$
(dotted), and ab-plane conductivity $\sigma_{ab} ( \omega )$ 
(dashed-dotted) as a function of frequency for 
$E_g=0.15t$ and $T=0.03t$ (arbitrary units). All
four quantities show gap-like suppressions at low frequencies.
The sizes of these gaps roughly have a relation of 1:2:3:4.
The ab-plane conductivity $\sigma_{ab} ( \omega )$ shows a strong
Drude peak at low frequencies within the gap.
}
\label{fig5}
\end{figure}

The in-plane conductivity $\sigma_{ab} ( \omega )$,
neglecting vertex corrections, is given by 
\begin{eqnarray}
\lefteqn{
\sigma_{ab} \left( \omega \right) = \frac{2 e^2}{\hbar c_0} 
\frac{\pi}{\omega} \int_{-\infty}^{\infty} d\omega'
\left[f(\omega') - f(\omega'+\omega)\right]
}\nonumber\\
& \times &
\frac{1}{N} \sum_{\bf k} \left[ v_{k,x}^2 + v_{k,y}^2\right]
\left[N({\bf k},\omega'+\omega)N({\bf k},\omega')
\right.
\label{eq11}
\\
& + &
\left.
A_1({\bf k},\omega'+\omega)A_1({\bf k},\omega') +
A_g({\bf k},\omega'+\omega)A_g({\bf k},\omega')
\right],
\nonumber
\end{eqnarray}
where $v_{k,i} = \partial \epsilon_k / \partial k_i$ are the
band velocities
within the ab-plane.
Our results for $\sigma_{ab}(\omega)$ (see Fig. 6 in
Ref. \onlinecite{DMTRaman}) are very similar to those for the
coherent c-axis conductivity shown in Fig. \ref{fig1}.
The in-plane conductivity is coherent in character
and shows a Drude peak at low frequencies even in the
underdoped compounds
\cite{Puchkov,Startseva}. However, the size of the pseudogap 
structure seen in ab-plane
conductivity on underdoped YBCO has been measured to be
600-700 cm$^{-1}$,
\cite{Puchkov,Basov} while the gap extracted from
c-axis conductivity in the
same compounds is of the order of
300-400 cm$^{-1}$. \cite{Homes,Uchida}
In Fig. \ref{fig5} we show our results for
$\sigma_{ab} ( \omega )$ and
$\sigma_c^{\rm incoh} ( \omega )$ for $E_g=0.15 t$
and $T=0.03 t$ along with
the density of states $N(\omega)$ Eq. (\ref{eq9}) and
the quasiparticle
damping rate $\omega {\rm Im} Z({\bf k_a},\omega)$ at the
antinodal momentum
${\bf k_a}$ at the Fermi surface (arbitrary units).
Here we see that the
size of the pseudogap appearing in these four quantities is quite
different. In fact, the gaps have an approximate relation
of 1:2:3:4
in the density of states, incoherent c-axis
conductivity, quasiparticle
damping rate, and ab-plane conductivity, respectively.
Especially, we
find that the gap structure in $\sigma_{ab} ( \omega )$
is about twice as 
big as in $\sigma_c^{\rm incoh} ( \omega )$, being about
$4\tilde{E}_g$ in
$\sigma_{ab} ( \omega )$ while only $2\tilde{E}_g$ in 
$\sigma_c^{\rm incoh} ( \omega )$ with
$\tilde{E}_g \approx 0.12t$, in rough 
agreement with experiment.
This relation of the gaps is a direct consequence of
the electronic
origin of the spinfluctuation scattering process and the
selfconsistency
of the FLEX equations: the opening of the pseudogap leads to a
suppression of the spinfluctuation interaction via
Eqs. (\ref{eq7}) and
(\ref{eq6}) at frequencies below $\sim 2\tilde{E}_g$. This in turn
results in a reduction of the self-energy below $\sim 3\tilde{E}_g$
and a corresponding structure at $\sim 4\tilde{E}_g$ in 
$\sigma_{ab} ( \omega )$. However, in the incoherent
c-axis conductivity
only a gap of size $\sim 2\tilde{E}_g$ appears because
of the momentum 
average of the spectral functions, resulting in a
frequency convolution
of the density of states with itself. The appearance
of a $4\Delta_0$-gap
in the ab-plane conductivity in the superconducting state for an
electronic pairing mechanism has been noted earlier in connection
with marginal Fermi liquid theory \cite{Varma,Nicol,DahmDiplom}.
Here we suggest that a corresponding effect is taking place in the
pseudogap state of underdoped high-$T_c$ compounds.

The pseudogap structures in the curves in Fig. \ref{fig5}
appear to be washed out somewhat and show more complex
behavior than a simple suppression at the effective
pseudogap. This is due to the fact that the pseudogap
Eq. (\ref{eq5}) is renormalized due to self-energy
effects. The structures seen in the conductivity,
density of states, and quasiparticle damping rate do not
display a pure $d$-wave gap, but a renormalized one,
similar as in the superconducting state (see Ref. 
\onlinecite{DahmTewordt}). More detailed discussions
of the density of states and the quasiparticle damping rate
can be found in Refs. \onlinecite{DahmTewordt,DMT2,DMTRaman}.

To summarize, we have investigated the influence of a normal-state
pseudogap of the form suggested by Williams et al \cite{Williams}
on the c-axis and ab-plane conductivity for spinfluctuation
exchange scattering within the selfconsistent FLEX approximation.
We find that coherent conductance can describe the
c-axis conductivity
in the overdoped compounds, while it is necessary to
consider incoherent
c-axis conductance in the underdoped regime. Incoherent conductance
can account well for the dynamical c-axis conductivity and the c-axis
resistivity in the underdoped compounds,
showing 'semiconducting' behavior.
However, it is difficult to reconcile the doping dependence of the
amplitude of the pseudogap, as suggested by the work of Williams et al,
with the doping independent size of the pseudogap seen in dynamical
c-axis and ab-plane conductivity. This suggests that the pseudogap
has a nontrivial momentum or frequency dependence, which changes
with temperature. We find that the difference in size of the pseudogap
in ab-plane conductivity as opposed to c-axis conductivity finds a 
natural explanation in the electronic origin of spinfluctuation
scattering and its selfconsistency with the single-particle properties.
This leads to a gap structure of size $\sim 4\tilde{E}_g$ in the 
ab-plane conductivity, while the gap seen in the incoherent
c-axis conductivity only has a size of $\sim 2\tilde{E}_g$.

\end{document}